# Fabricating Highly Ordered Nanofiber Assemblies by Controlled Shear Flow and Solvent Evaporation

Enlai Gao[1], Chuanhua Duan[2,*] and Zhiping Xu[1,*]

[1]Applied Mechanics Laboratory, Department of Engineering Mechanics, and Center for Nano and Micro Mechanics, Tsinghua University, Beijing 100084, China

[2]Department of Mechanical Engineering, Boston University, Boston, MA 02215, USA

*Email: xuzp@tsinghua.edu.cn, duan@bu.edu

## Abstract

Fabrication of highly ordered and dense nanofibers assemblies is of key importance in designing high-performance and multi-functional materials. In this work, we design an experimental approach *in silico*, combining shear flow and solvent evaporation to establish nanofiber alignment and densification of the assemblies. We demonstrate the feasibility of this approach by performing dissipative particle dynamics simulations, where the hydrodynamic and thermal fluctuation effects are fully modeled. We find that the microstructural order of assembled nanofibers can be established within a specific range of the Peclet number and evaporation flux. The underlying mechanism is elucidated by considering the competition between hydrodynamic coupling and fluctuating dynamics of nanofibers. Based on these understandings, we outline a practical setup to fabricate highly ordered and dense nanofiber assemblies.

# 1. Introduction

Nanostructures such as nanowires, nanotubes and nanosheets can be utilized as elementary building blocks in constructing high-performance, multi-functional materials and devices [1, 10, 37]. The advantages of this bottom-up approach include function-oriented structural tailoring and engineering from the molecular and microstructural levels up to macroscopic scales where materials are usually used. However, to harness unconventional attributes of nanostructures to a macroscopically ordered material is a highly challenging problem. A significant number of theoretical models have been proposed to elucidate the correlation between their microstructures and macroscopic performance [5, 14, 18, 19, 40]. Rationally optimized material is thus straightforward in this new dimension of design, but experimental advances along this direction are retarded due to the lack of precise and efficient control of the microstructural organization in the assemblies of nanostructures. Insights from theoretical investigation have been offered to understand these processes. For example, it is worth noting that although directed self-assembly could be tuned to obtain nanofiber networks with various topology, ranging from lattice structures to liquid-crystal-like microstructures, the lack of directional driven force cannot assure the production of dense and ordered microstructures for high material performance. As a result, sparse networks with considerable local topological defects of nanofiber assemblies form during their directed self-assembly process [35].

One powerful way to assemble nanofibers with outstanding mechanical properties is the process inspired by spider silk spinning in Mother Nature, with dry, wet, melting and other spinning techniques as its industrial counterparts [17, 28]. Fibers, sheets or other building blocks are suspended in solution for dispersion, and then exposed under uniaxial viscous flow where solvents are evaporated to improve the microstructural ordering [32, 38]. Spinning and filtration techniques have also been widely adopted to assemble fibers and membranes from nanofibers and nanosheets, which are promising as new classes of engineering materials with enhanced mechanical, thermal, electrical and optical properties [9, 13, 30]. However, the quality of assembled structures and understanding of the underlying mechanisms at the microstructural level need to be further improved [19, 27, 38]. The processes of microstructural ordering in the fluid phase and fluid-to-solid transition have to be well understood and controlled to fabricate macroscopic assemblies with improved microstructural alignment and densification.

For flow-driven dynamics of microstructures, studies have been carried out to understand the orientational dynamics of ellipsoid and cylinders. Jeffrey [11] developed theories that predict the behaviors of a rigid ellipsoid or cylinder exposed to hydrodynamic forces only, in the absence of Brownian motion [23]. The ellipsoid undergoes an orbital motion in viscous flow, and the trajectory is known later as the Jeffery orbit. Computer simulations and experiments were conducted to explore these orbits by considering the Brownian motion of a single cylinder, which suggest that orientation of the cylinder prefers to be spatially homogeneous due to thermal fluctuation [4, 15, 29, 31]. Tannous [33] showed that the two-dimensional probability density functions (PDF) of cylinders satisfies the Fokker-Planck equation, and the driven flow improves the orientational order of the assemblies. Rigid solid objects align to the direction of shear flow and the orientation order can be expressed as function of the Peclet number, $Pe = \gamma/D_r$, where $\gamma$ is the shear rate and $D_r$ is the rotational diffusion coefficient. Moreover, in a recent experimental study, the relaxation of flow-induced orientation of multiwalled carbon nanotubes (CNTs) was observed following flow cessation [26], although detailed correlation between the flow-structure interaction (*e.g.* fluid properties, flow rate, nanofiber concentration, shape and interaction) and the assembled structures still awaits clarification.

In addition to fiber alignment, densification is another key figure of merit in the assembly process for high-performance materials. Solvent evaporation is one general route to make the nanofibers well-aligned and densely packed [3, 39]. Marin and his collaborators [21, 22] demonstrated an order-to-disorder transition of particles in coffee-ring stains as an evaporation-induced assembly process. Bigioni *et al.* [2, 24] showed that relative time scales of particle accumulation and diffusion determine the assembly behavior and growth kinetics of monolayer gold nano-particles. Experiments were also conducted by introducing evaporation in assembling superstructures [3, 16]. However, the synergistic effect of shear flow and evaporation on the highly ordered and dense assembly structure is not well discussed.

Our understanding of how to precisely assemble highly ordered and dense assembly structures through shear flow and solvent evaporation need thus to be advanced, for example, with knowledge and insights from their microscopic dynamics. In this work, we present a numerical study based on dissipative particle dynamics (DPD) simulations for the flow and evaporation-induced assembly of nanofibers in liquid. We consider factors that influence the assembled structure such as the shear rate of flow, fluid properties, shape of the nanofibers, nanofiber-nanofiber, nanofiber-fluid interaction, and the evaporation kinetics. We identify the viable conditions in fabricating highly ordered and dense nanofiber assemblies and propose a practical fluidic channel design to fabricate highly ordered and dense nanofiber assemblies. The rest of this paper is organized as follows. The models and methods are described in Sec. 2. DPD simulation results, theoretical analysis and the design of fluidic channel are presented and discussed in Sec. 3, before the conclusion is made in Sec. 4.

## 2. Models and methods

In this work, we simulate the flow and evaporation-driven assembly of nanofibers using two-dimensional (2D) DPD, which is a mesoscale particle-based coarse-graining simulation method [6, 7]. The motion of the DPD particles is governed by Newton's equations, and the force on particle can be decomposed into three parts of contributions

$$F_i = \Sigma_{j \neq i}(F_{ij}^C + F_{ij}^D + F_{ij}^R) \tag{1}$$

Here $C$, $D$, and $R$ stand for conservative, dissipative, and random, respectively. For a bead in the single-component DPD, the force exerted on particle $i$ by particle $j$ includes

$$F_{ij}^C = a(1 - r_{ij}/R_c)\hat{\mathbf{r}}_{ij}, \; r_{ij} < R_c \tag{2}$$

$$F_{ij}^D = -\gamma w^D(r_{ij})(\hat{\mathbf{r}}_{ij} \cdot \mathbf{v}_{ij})\hat{\mathbf{r}}_{ij} \tag{3}$$

$$F_{ij}^R = \sigma_R w^R(r_{ij})\theta_{ij}\hat{\mathbf{r}}_{ij} \tag{4}$$

Here $\mathbf{v}_{ij} = \mathbf{v}_i - \mathbf{v}_j$, $\mathbf{r}_{ij} = \mathbf{r}_i - \mathbf{r}_j$, $r_{ij} = |\mathbf{r}_{ij}|$ and $\hat{\mathbf{r}}_{ij} = \mathbf{r}_{ij}/r_{ij}$. The conservative force $F_{ij}^C$ is soft repulsion acting along the line of bead centers and $a = a_{ij} = (a_i a_j)^{1/2}$, where $a_i$ and $a_j$ are conservative force coefficients for particles $i$ and $j$, and $R_c$ is the cut-off distance for this soft repulsion. The coefficients $\gamma$ in $F_{ij}^D$ and $\sigma_R$ in $F_{ij}^R$ define the amplitudes of the dissipative and random forces, respectively. $w^D$ and $w^R$ are weight functions that vanish

for $r_{ij} > R_c$, and $\theta_{ij}$ is a Gaussian random variable that satisfies $<\theta_{ij}(t)> = 0$ and $<\theta_{ij}(t)\theta_{kl}(t')> = (\delta_{ik}\delta_{jl} + \delta_{il}\delta_{jk})\delta(t - t')$. To ensure that the Boltzmann distribution is achieved at thermal equilibrium, we require $w^D(r_{ij}) = [w^R(r_{ij})]^2$ and $\sigma_R^2 = 2\gamma k_B T$. For simplicity, $w^D(r_{ij})$ is chosen to be $1 - r_{ij}/R_c$, which defines the same interaction region as that for the conservative force.

We coarse grain both solvent and nanofibers in the DPD simulations. For the solvent, nine water molecules are grouped into one bead. The molecular volume of water is $v_w = 30$ Å$^3$, so that one bead represents the volume of $V_w = 270$ Å$^3$. The number density of DPD beads is chosen to be $\rho R_c^3 = 3$, where $\rho$ is the number of DPD beads per cubic volume $R_c^3$. Accordingly, we have $R_c = 9.32$ Å and $m = 2.69 \times 10^{-25}$ kg. The DPD time scale for the assembly dynamics, $\tau = 549$ ps, is obtained by matching the bulk diffusion constant of water [6]. In the DPD simulations, we use $R_c$, $k_B T$, $m$ and $\tau$ as the reference units for length, energy, mass and time [20]. The value of parameter $a$ is fitted to reproduce the solvent compressibility as shown in **Fig. S1** in the **Supplementary Material**. In summary, these settings are validated by reproducing the mass density, 1000 kg/m$^3$, diffusion coefficient, $2.43 \pm 0.01 \times 10^{-5}$ cm$^2$/s, and compressibility, 2.2 GPa, of water, as summarized in **Table S1** in the **Supplementary Material**.

For nanofibers, without loss of generality, we use material parameters for the multi-walled CNT as adopted in our previous work [35]. Within each CNT, the stretching contribution between two bonding beads to the total energy is given by $E_T = K_T(R - R_0)^2/2$, where $K_T = YA/2R_0$ is the spring constant related to the tensile stiffness, expressed in the Young's modulus $Y$, the cross-section area $A$, and distance between two bonding beads $R$. $R_0$ is the equilibrium inter-bead distance. The bending energy contribution within an adjacent beads triplet is described as $E_B = K_B(1 + \cos\theta)$, where $K_B = 3YI/2R_0$ is the angular spring constant related to the bending stiffness $D = YI$, $\theta$ is bending angle within the triplet, and $I$ is the bending moment of inertia. The interfiber ineraction is in the soft repulsion form (**Eq. 2**) where fiber crossing is prohibited. These force field parameters are summarized in **Table S2** in the **Supplementary Material**.

All DPD simulations run under the constant-temperature condition at 300 K using the DPD thermostat, which conserves the total momentum of system and correctly captures the hydrodynamics [6, 7]. To maintain the balance between numerical stability and the computational consumption, we use two different time step sizes for the DPD beads of solvent (~$10^5$ beads) and nanofibers (~$10^3$ beads), which are $0.005\tau$ and $0.0001\tau$, respectively, implemented through the reversible reference system propagator algorithms (rRESPA) multi-timescale integrator with two hierarchical levels [34]. More details can be found in the **Supplementary Material**. Equilibrium molecular dynamics (EMD) simulations are performed to measure the rotational coefficient based on the Einstein relation, by tracking the angular mean square displacement (AMSD). The Lees-Edwards's boundaries conditions are applied to generate the shear flow, so that the same shear rate are applied to the whole system globally. Periodic boundary conditions (PBCs) are applied to the simulation box with a fixed number of nanofibers dispersed in the dilute solution regime, with $cl^2 < 1$ in 2D. Here $c$ is the number density of nanofibers and $l$ is the mean length of nanofibers [8]. All simulations are performed using the large scale atomic/molecular massively parallel simulator (LAMMPS) [25].

## 3. Results and discussion

The process of flow-driven assembly is illustrated in **Fig. 1**, where three simulation snapshots summarize the key steps in forming the assemblies. The first step is to disperse the nanofibers in solution with randomly distributed position and orientation, while the second step is to expose them under controlled shear flow, which introduces a high orientational order within the nanofiber assemblies. The third step is to evaporate the solvent to reach the final assemblies with highly ordered and dense microstructures. To gain more insights into these processes, we analyze the controlling steps of shear flow and evaporation in the assembly process.

We define an orientation order parameter (OOP) $O$ to quantify the orientational order of nanofibers in the assembly, which is also called nematic order parameter since it is zero for the randomly aligned isotropic phase and one for the perfectly aligned phase, respectively. Consequently, the alignment of nanofibers to the flow direction ($\theta = 0$) is signaled by the saturation of $O$ towards 1, since in 2D we have

$$O = \langle 2\cos^2\theta - 1 \rangle \tag{5}$$

where definition of $\theta$ is shown in **Fig. 1c** and **2b**. In the assembly dynamics, orientation of the nanofibers is determined by the competition between hydrodynamic forces and Brownian random perturbation. The Peclet number $Pe$ measures the relative strength of hydrodynamic forces and Brownian forces, that is

$Pe = \gamma/D_r$ (6)

Here the rotational coefficient $D_r$ is measured through the tracked AMSD in equilibrating DPD simulations using the Einstein relation in 2D, *i.e.*

$D_r = \Delta\theta^2/2t$ (7)

To probe the effect of Brownian forces while keeping other factors fixed, we first consider various lengths of nanofibers, $l$, in our simulations. The rotational trajectory of nanofibers in equilibration simulations is plotted in **Fig. S4a**, with the AMSD in **Fig. S4b** in the **Supplementary Material**. These results demonstrate the distinct nature of Brownian rotational motion, and that the rotational diffusion coefficient $D_r$ decreases with the increase in the length of nanofibers.

We apply a shear rate of $\gamma = 0.01\tau^{-1}$ to the solvent where these nanofibers are dispersed and monitor the PDF of their orientational distribution, as well as the value of $O$. From the results summarized in **Fig. 2**, one find that the PDF of nanofiber orientation distribution becomes sharply peaked around the most probable orientation angle, indicating a certain degree of orientational order. The parameter $O$ plotted in **Fig. 4a** demonstrates that OOP increases, while $D_r$ decreases as $l$ increases. A similar trend was reported in the previous study of CNTs in argon gas shear flow [4].

We then explore the role of hydrodynamic forces exclusively, by changing the shear rate $\gamma$ while keeping other factors fixed. The distribution of nanofibers becomes sharply peaked around the most probable orientation and OOP increase with the shear rate as shown in **Figs. 3** and **4b** in low Pelect number region. However, this trend reverses as the Pelect number exceeds a critical value of $Pe = \sim 95$. This finding at low $Pe$ values agrees with the analytical and numerical studies by Tannous [33], where inertial effects are excluded. In contrast, the observation in high $Pe$ number region is anomalous as shown in **Fig. 4b**, and should be understood as an outcome of the inertial effect at high shear rates. The homogenous distribution of nanofibers resulted from the inertial effect can be measured as Reynolds number *Re*, calculated from the shear rate and the length of nanofiber as

$$Re = \gamma \rho (l/2)^2 / \mu \tag{8}$$

where $\rho$ and $\mu$ are the density and dynamic viscosity of the fluid. In addition, we have to point that the Reynolds number should not exceed $10^3 \sim 10^4$ for laminar flow, as flow with very high shear rate not only induces a anomalous distribution of nanofibers as shown in **Fig. 4b**, but also leads to a transition from laminar flow to turbulence. Therefore, this anomalous region of Peclet number (or shear rate) should be avoided in practical design, and our following discussions in this work only focus on the region of Peclet number that has a positive correlation between the Peclet number and OOP.

Highly ordered nanofiber assemblies can be achieved in a specific range of Pelect number. However, experimental study has reported the relaxation of orientational order following shear flow cessation [26]. We confirm this result in our DPD simulation results, which are summarized in **Fig. 5**. Our DPD simulation results show that, in the absence of evaporation (**Fig. 5a**), the OOP increases, reaches a constant under shear flow at *t* from 0 to $5\times10^3\tau$, and then relaxes to zero after shear flow cessation after $t = 5\times10^3\tau$. While in the condition that 85% of the solvent is evaporated (**Fig. 5b**, simulation details can be found in the **Supplementary Material**), the orientational order is partially preserved after the shear flow cessation. In addition, from the DPD simulation results, we find that the density of the oriented structure of nanofibers from dilute solution region is too low to fabricate high-performance materials in fiber or film forms. Although low density of nanofibers reduces the local topological defects in the microstructures during the assembly process [35], it prevents wide applications because of the poor mechanical, thermal and electrical properties of the assembly. Hence the density and microstructural stability of nanofiber assemblies need to be improved, and solvent evaporation becomes a feasible approach. If evaporation is introduced, the orientational order can be partially preserved even after flow cessation. By exploring the effect of evaporation fluxes, we conclude from the simulation results that solvent evaporation leads to ordered and stable assemblies as shown in **Fig. 5**.

From our previous analysis, highly ordered and dense assembled structures of nanofiber assemblies can be obtained by controlling the flow and evaporation. Based on these understandings, we propose a design of a fluidic channel where the nature of flow and evaporation can be engineered.

The profile of 2D reducing fluidic channel can be described with two profiles as $h(x)$ and $-h(x)$, as shown in **Fig. 6a**. The width, pressure and flux of inlet (outlet) are $h_1$ ($h_2 = \beta h_1$, where $\beta = h_2/h_1$ defines the channel width reduction), $p_1$ ($p_2$) and $q_1$ ($q_2$), respectively, and the length of channel is $L$. From the Navier-Stokes equation for one dimensional and incompressible flow that disregards gravity and inertia forces for the assembly of nanostructures, we have

$$\frac{d^2 u(x,y)}{dy^2} = \frac{1}{\mu} \frac{dp(x)}{dx} \tag{9}$$

where $u$ and $p$ are the flow velocity and pressure along channel, respectively. This equation can be solved with the boundary conditions $u(x, y = \pm h(x)) = 0$ and the equation of continuity, which gives the spatial distribution of flow velocity, pressure, and the correlation between flux $q$ and pressure difference $\Delta p = p_1 - p_2$. Detailed derivation can be found in the **Supplementary Material**.

$$u(x,y) = \frac{\Delta p}{\mu L} \frac{h_1^2 h_2^2}{(h_1 + h_2) h(x)^3} [h(x)^2 - y^2] \tag{10}$$

$$p(x) = p_1 - \frac{[h_1/h(x)]^2 - 1}{(h_1/h_2)^2 - 1} \Delta p \tag{11}$$

$$q(x) = \frac{4 d h_1^2 h_2^2}{3\mu L (h_1 + h_2)} \Delta p = q_1 = q_2 \tag{12}$$

Here $d$ is the depth of this 2D channel. The shear rate distribution in the channel can then be calculated from the $u(x, y)$, which strongly affects the orientational order, while pressure difference $\Delta p$ and flux $q$ could be used in other design steps such as the energy consumption and structural stability, as shown in **Figs. S2** and **S3**.

To further consider the effect of evaporation, a supplementary condition is introduced that satisfies the mass conservation of solvent. With the assumption that total evaporation flux ($q_e = \lambda A$) is proportional to the flow area swept from inlet to $x$ along the channel $A = [h_1^2 - h(x)^2]/\tan\alpha$, the coefficient $\lambda$ defines the intensity of evaporation. Hence, the flux of fluid $q(x)$ at position $x$ can then be expressed as

$$q(x) = q_1 - q_e(x) = q_1 - \lambda \frac{h_1^2 - h(x)^2}{\tan a} \tag{13}$$

The distribution of average shear rate along the channel significantly affects the orientational order of nanofibers in the flow, which can be simply defined as

$$\gamma(x) = \frac{q(x)}{dh(x)^2} \tag{14}$$

By combining **Eqs. 13** and **14**, we solve $\gamma(x)$ and plot the results for $\gamma(x)$ in **Fig. 6b**, where the dash and solid lines indicate the evaporation-absent condition ($q_2 = q_1$) and strong evaporation limit where $q_2$ at the outlet vanishes ($q_2 \approx 0$, considering the volume of nanofibers is far lower than that of the solvent at inlet, as the all solvent evaporates along the channel, the outlet flux almost vanishes), respectively. On the evaporation-absent condition, the average shear rate increases along the channel from the inlet to outlet, due to the gradual reduction of channel width $h$ as shown in **Fig. 6a**. While in the strong evaporation limit, the trend of shear rate distribution along the channel is reversed. As shown before, the orientational order of nanofibers will relax under shear flow cessation or shear rate reduction, so the shear rate at the outlet plays an important role in determining the the orientational order of nanofiber assemblies. Furthermore, our discussions focus on the region of *Pe* number with positive correlation between the *Pe* number and OOP. Hence, the strong evaporation limit does not help to achieve higher orientational order in practice due to the reducing shear rate distribution from inlet to outlet, and should be prevented in experimental design.

In order to rationally control the orientational order and densification of the nanofiber assemblies, we predict the correlation between the shear rate for orientational ordering, density of nanofibers at the outlet and the shape of fluidic channel (the inlet, outlet width and the length of channel), evaporation intensity coefficient $\lambda$. The shear rate at outlet $\gamma_{out}$ that controls the orientational order of product assemblies can be derived from **Eqs. 13** and **14**

$$\gamma_{out} = \frac{q_1 - \lambda(h_1 + h_2)L}{dh_2^2} \tag{15}$$

Here $\lambda(h_1+h_2)L$ is the evaporation flux from the inlet to outlet. The density of nanofiber assemblies at the outlet $\rho_{out}$ can be related to the evaporation flux $\lambda(h_1+h_2)L$ and inlet flux $q_1$. Accordingly, the density of nanofiber assemblies at the outlet $\rho_{out}$ can be defined as

$$\rho_{out} \sim 1/[\, q_1 - \lambda(h_1+h_2)L] \tag{16}$$

This indicates that the higher flux the solvent evaporates in the flow process, the higher density of assembled structure it could reach. The results of **Eqs. 15** and **16** are shown in **Fig. 7a**, from which we can see that, the shear rate and density of nanofiber assemblies at the outlet can be controlled through adjusting the shape of fluidic channel. In general, longer channel and smaller width of outlet, compared to the width of inlet, is beneficial to fabricate highly ordered and dense assemblies of nanofibers. In addition, from **Eqs. 15**, **16** and **Fig. 7b**, both the shear rate and density of nanofiber assemblies at the outlet depend on the evaporation intensity coefficient $\lambda$, which suggests that the orientational order and density of assemblies can be adjusted through the evaporation intensity coefficient $\lambda$. To summarize, **Fig. 7** illustrates the strategy to access specified density and orientational order of nanofiber assemblies through adjusting the shape of fluidic channel, evaporation intensity and applied pressure gradient or flux.

## 4. Conclusion

In this study, we conducted DPD simulations to explore the dynamic processes of flow and evaporation-induced assembly of nanofibers. Admittedly the results presented here are limited as we consider only identical nanofibers with the same geometry and all the simulations and discussion are carried out in 2D. However, we corroborate recent findings in the experiments of controllable assembly of nanofibers [12, 16, 21, 26, 36, 38] and explore the key parameters in the process that determine the microstructure of final assembly. The major findings of this work include that the orientational order and density of assembled structure can be well engineered through parametric design of the experimental setup. Moreover, the shear flow and evaporation processes have a synergistic effect on the microstructural formation. This study further highlight the concept that nanostructures such as nanotubes, nanowires, nanosheets, cyclic peptides can be utilized as fundamental building blocks in constructing hierarchical materials and structures, and provide key insights into practical solution for this goal in a fluid-based approach.

## ACKNOWLEDGMENTS

This work was supported by the National Natural Science Foundation of China through Grant 11222217 and 11472150, and the State Key Laboratory of Mechanics and Control of Mechanical Structures (Nanjing University of Aeronautics and Astronautics) through Grant No. MCMS-0414G01. The simulations were performed on the Explorer 100 cluster system of Tsinghua National Laboratory for Information Science and Technology.

# FIGURES AND FIGURE CAPTIONS

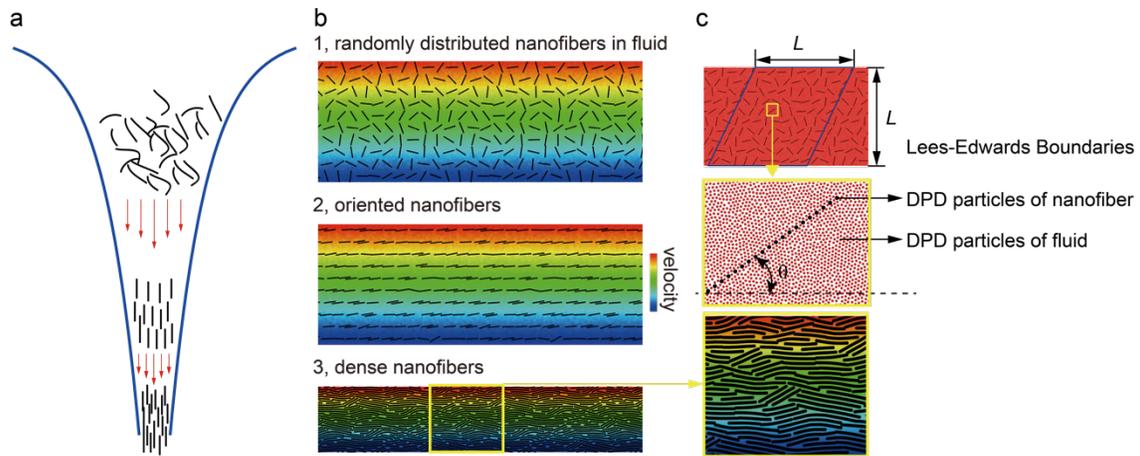

**Figure 1** (a) A schematic illustration of the flow-evaporation process for nanofiber assembly. (b) DPD simulation snapshots show the alignment and densification of nanofibers. Only nanofibers are plotted, and the color map indicates the amplitude of velocity field in fluid. (c) Detailed snapshots show the Lees-Edwards boundary condition to simulate the shear flow and DPD particles for nanofibers and fluid.

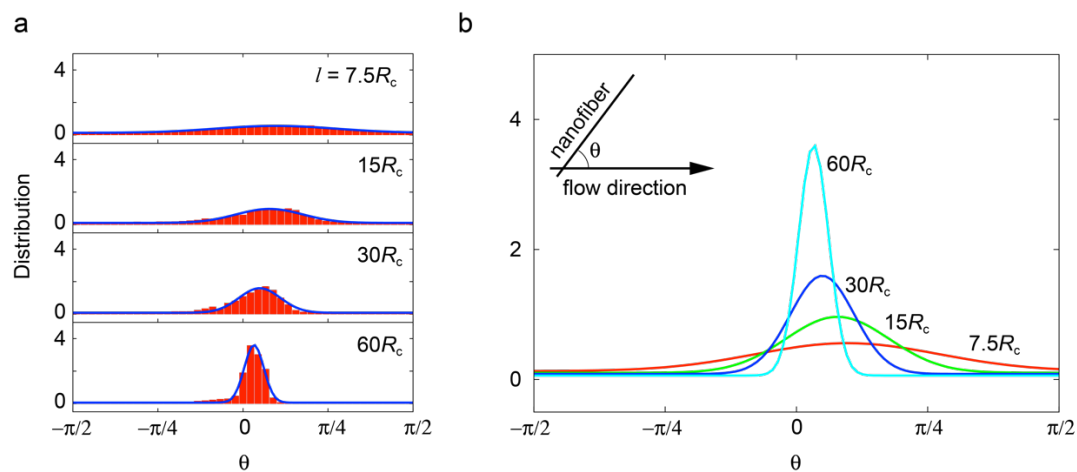

**Figure 2** (a) Probability distribution functions (PDFs) of nanofiber orientation. The red boxes are simulation results for different lengths of nanofibers *l*, fitted to the Gaussian distribution plotted in solid blue lines. (b) Comparison of PDFs shows the increase of the peak value and narrowing of the width as *l* increases, indicating improved alignment of nanofibers.

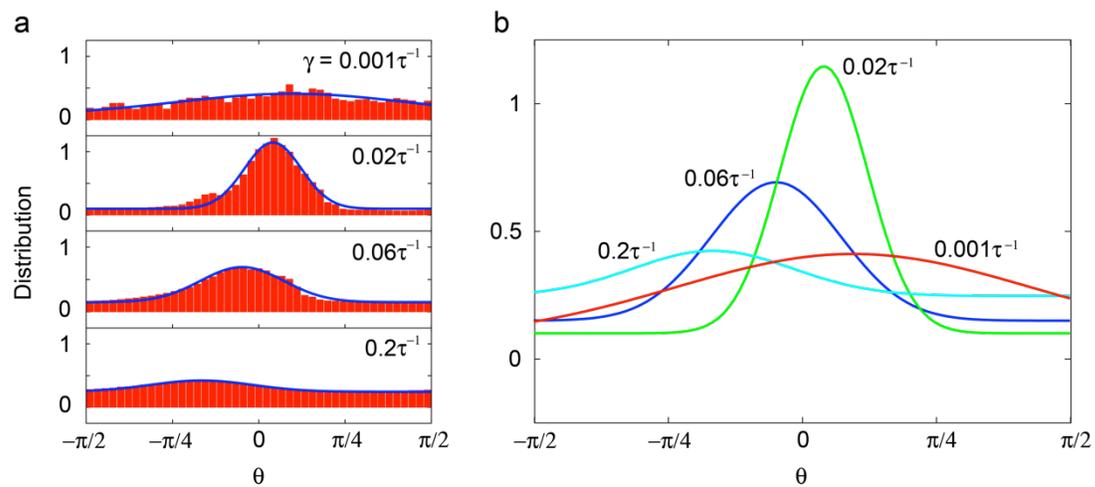

**Figure 3** (a) PDFs of nanofiber orientations, plotted for different shear rates $\gamma$, which are also fitted to the Gaussian distribution. (b) Comparison between the PDFs shows the shift of the peak value and change in the width as $\gamma$ increases.

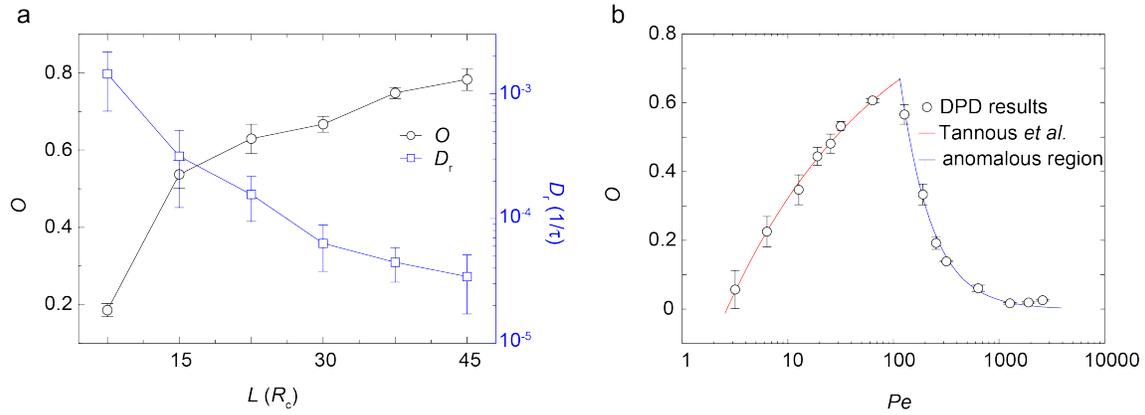

**Figure 4** (a) The correlation between the rotational diffusion coefficient $D_r$, orientational order paramenter $O$, and the length of nanofiber $l$. Our simulation results show that $D_r$ decreases with $l$, resulting in a higher Pelect number and increase in $O$. (b) The correlation between $O$ and the shear rate $\gamma$. The value of orientational order parameter $O$ increases first with $\gamma$ athe the low-$Pe$ regime but decreases after $Pe$ exceeds a critical value of ~95.

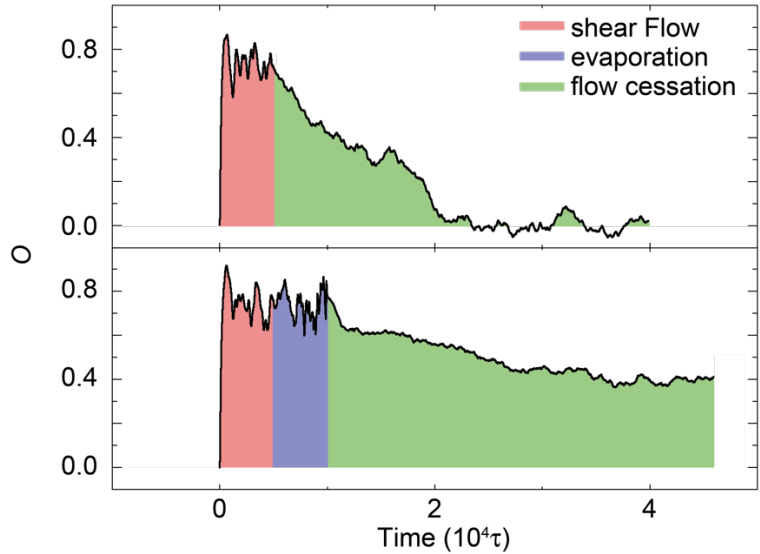

**Figure 5** Relaxation of the orientational order in nanofiber assemblies after flow censsation. (a) In the absence of evaporation, the OOP increases, reaching a constant under shear flow ($t = 0 - 5\times10^3\ \tau$), and then relaxes to zero after shear flow cessation (after $5\times10^3\ \tau$). (b) With evaporation, the OOP increases and becomes saturated soon. $O$ keeps almost as a constant in $\sim 1\times10^4\tau$, where evaporation (85% of the solvent) is introduced from $t = 5\times10^3\tau$ to $1\times10^4\tau$. After the shear flow cessation from $\sim 1\times10^4\tau$, $O$ relaxes to $\sim 0.4$ and and the orientational order is partially preserved. The light-red, light-blue and light-green regions are the stages of steady shear flow, solvent evaporation and flow cessation, respectively.

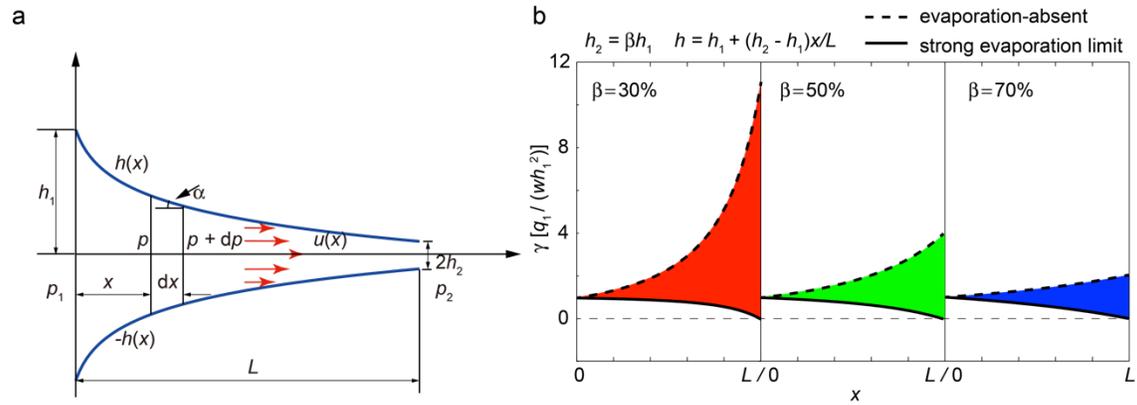

**Figure 6** (a) The profile of fluidic channel. (b) Average shear rate distribution along the fluidic channel for different geometrical parameters $\beta = h_2/h_1$. The dash and solid lines show the predictions for the evaporation-absent condition and strong evaporation limit, respectively.

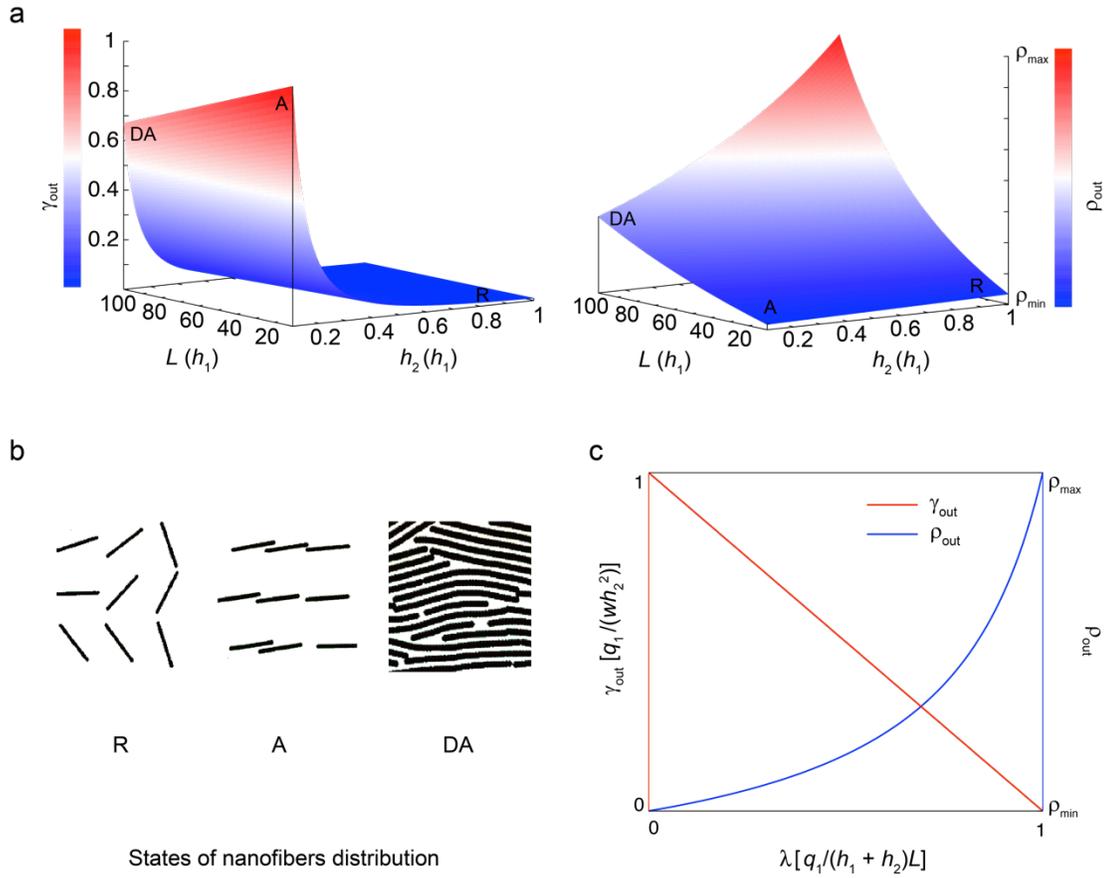

**Figure 7** (a) The dependence of outlet shear rate $\gamma_{out}$, density $\rho_{out}$ on the shape of fluidic channel, expressed in the ratios of channel length and outlet width to the inlet width, *i.e.* $L/h_1$ and $h_2/h_1$. The value of $\gamma_{out}$ is normalized by its maxima. (b) R, A, and DA stand for the assembly with random, aligned, aligned and densified microstructures obtained from DPD simulations, respectively. (b) The dependence of $\gamma_{out}$ and $\rho_{out}$ on the evaporation intensity coefficient $\lambda$.

# REFERENCES

[1] Behabtu N, Young C C, Tsentalovich D E, Kleinerman O, Wang X, Ma A W K, Bengio E A, ter Waarbeek R F, de Jong J J, Hoogerwerf R E, Fairchild S B, Ferguson J B, Maruyama B, Kono J, Talmon Y, Cohen Y, Otto M J and Pasquali M 2013 Strong, Light, Multifunctional Fibers of Carbon Nanotubes with Ultrahigh Conductivity *Science* **339** 182-6

[2] Bigioni T P, Lin X M, Nguyen T T, Corwin E I, Witten T A and Jaeger H M 2006 Kinetically driven self assembly of highly ordered nanoparticle monolayers *Nat. Mater.* **5** 265-70

[3] Brinker C J, Lu Y F, Sellinger A and Fan H Y 1999 Evaporation-induced self-assembly: Nanostructures made easy *Adv. Mater.* **11** 579

[4] Dong R Y and Cao B Y 2014 Anomalous orientations of a rigid carbon nanotube in a sheared fluid *Sci. Rep.* **4** 6120

[5] Egan P, Sinko R, LeDuc P R and Keten S 2015 The role of mechanics in biological and bio-inspired systems *Nat. Commun.* **6** 7418

[6] Groot R D and Rabone K L 2001 Mesoscopic simulation of cell membrane damage, morphology change and rupture by nonionic surfactants *Biophys. J.* **81** 725-36

[7] Groot R D and Warren P B 1997 Dissipative particle dynamics: Bridging the gap between atomistic and mesoscopic simulation *J. Chem. Phys.* **107** 4423-35

[8] Hobbie E K 2010 Shear rheology of carbon nanotube suspensions *Rheol. Acta* **49** 323-34

[9] Hobbie E K and Fry D J 2006 Nonequilibrium phase diagram of sticky nanotube suspensions *Phys. Rev. Lett.* **97** 036101

[10] Huang H, Song Z, Wei N, Shi L, Mao Y, Ying Y, Sun L, Xu Z and Peng X 2013 Ultrafast viscous water flow through nanostrand-channelled graphene oxide membranes *Nat. Commun.* **4** 3979

[11] Jeffery G B 1922 The Motion of Ellipsoidal Particles Immersed in a Viscous Fluid *Proc. Roy. Soc. A* **102** 161-79

[12] Ko H and Tsukruk V V 2006 Liquid-crystalline processing of highly oriented carbon nanotube arrays for thin-film transistors *Nano Lett.* **6** 1443-8

[13] Kwon G, Heo Y, Shin K and Sung B J 2012 Electrical percolation networks of carbon nanotubes in a shear flow *Phys. Rev. E* **85** 011143

[14] Launey M E, Buehler M J and Ritchie R O 2010 On the Mechanistic Origins of Toughness in Bone *Annu. Rev. Mater. Res.* **40** 25-53

[15] Leal L G and Hinch E J 2006 The effect of weak Brownian rotations on particles in shear flow *J. Fluid Mech.* **46** 685

[16] Lee S G, Kim H, Choi H H, Bong H, Park Y D, Lee W H and Cho K 2013 Evaporation-induced self-alignment and transfer of semiconductor nanowires by wrinkled elastomeric templates *Adv. Mater.* **25** 2162-6

[17]   Lin S, Ryu S, Tokareva O, Gronau G, Jacobsen M M, Huang W, Rizzo D J, Li D, Staii C, Pugno N M, Wong J Y, Kaplan D L and Buehler M J 2015 Predictive modelling-based design and experiments for synthesis and spinning of bioinspired silk fibres *Nat. Commun.* **6** 6892

[18]   Liu Y, Xie B, Zhang Z, Zheng Q and Xu Z 2012 Mechanical Properties of Graphene Papers *J. Mech. Phys. Solids* **60** 591-605

[19]   Lu W, Zu M, Byun J H, Kim B S and Chou T W 2012 State of the art of carbon nanotube fibers: opportunities and challenges *Adv. Mater.* **24** 1805-33

[20]   M. P. Allen D J T 1987 *Computer Simulation of Liquids* (Clarendon Press)

[21]   Marin A G, Gelderblom H, Lohse D and Snoeijer J H 2011 Order-to-disorder transition in ring-shaped colloidal stains *Phys. Rev. Lett.* **107** 085502

[22]   Marín A l G, Gelderblom H, Lohse D and Snoeijer J H 2011 Rush-hour in evaporating coffee drops *Phys. Fluids* **23** 091111

[23]   Meirson G and Hrymak A N 2015 Two-dimensional long-flexible fiber simulation in simple shear flow *Polym. Compos.* (published online, doi: 10.1002/pc.23427)

[24]   Narayanan S, Wang J and Lin X M 2004 Dynamical self-assembly of nanocrystal superlattices during colloidal droplet evaporation by in situ small angle x-ray scattering *Phys. Rev. Lett.* **93** 135503

[25]   Plimpton S 1995 Fast Parallel Algorithms for Short-Range Molecular-Dynamics *J. Comp. Phys.* **117** 1-19

[26]   Pujari S, Rahatekar S S, Gilman J W, Koziol K K, Windle A H and Burghardt W R 2009 Orientation dynamics in multiwalled carbon nanotube dispersions under shear flow *J. Chem. Phys.* **130** 214903

[27]   Richardson J J, Bjornmalm M and Caruso F 2015 Multilayer assembly. Technology-driven layer-by-layer assembly of nanofilms *Science* **348** 2491

[28]   Rising A and Johansson J 2015 Toward spinning artificial spider silk *Nat. Chem. Biol.* **11** 309-15

[29]   Sader J E, Pepperell C J and Dunstan D E 2008 Measurement of the optical properties and shape of nanoparticles in solution using Couette flow *ACS Nano* **2** 334-40

[30]   Shim J S, Yun Y H, Rust M J, Do J, Shanov V, Schulz M J and Ahn C H 2009 The precise self-assembly of individual carbon nanotubes using magnetic capturing and fluidic alignment *Nanotechnology* **20** 325607

[31]   Stover C A, Koch D L and Cohen C 2006 Observations of fibre orientation in simple shear flow of semi-dilute suspensions *J. Fluid Mech.* **238** 277

[32]   Su B, Wu Y and Jiang L 2012 The art of aligning one-dimensional (1D) nanostructures *Chem. Soc. Rev.* **41** 7832-56

[33]   Tannous C 2011 Langevin simulations of rod-shaped object alignment by surface flow *Surf. Sci.* **605** 923-9

[34]   Tuckerman M, Berne B J and Martyna G J 1992 Reversible Multiple Time Scale

Molecular-Dynamics *J. Chem. Phys.* **97** 1990-2001

[35] Xie B, Buehler M J and Xu Z 2015 Directed self-assembly of end-functionalized nanofibers: from percolated networks to liquid crystal-like phases *Nanotechnology* **26** 205602

[36] Xie D, Lista M, Qiao G G and Dunstan D E 2015 Shear Induced Alignment of Low Aspect Ratio Gold Nanorods in Newtonian Fluids *J. Phys. Chem. Lett.* **6** 3815-20

[37] Xu Z and Buehler M J 2009 Hierarchical nanostructures are crucial to mitigate ultrasmall thermal point loads *Nano Lett.* **9** 2065-72

[38] Xu Z and Gao C 2014 Graphene in macroscopic order: liquid crystals and wet-spun fibers *Acc. Chem. Res.* **47** 1267-76

[39] Zhang M, Wang Y, Huang L, Xu Z, Li C and Shi G 2015 Multifunctional Pristine Chemically Modified Graphene Films as Strong as Stainless Steel *Adv. Mater.* **27** 6708-13

[40] Zhu H, Zhu S, Jia Z, Parvinian S, Li Y, Vaaland O, Hu L and Li T 2015 Anomalous scaling law of strength and toughness of cellulose nanopaper *Proc. Natl. Acad. Sci.* **112** 8971-6